\pdfoutput=1
\documentclass[10pt]{article}
\usepackage{graphicx}
\usepackage{color}
\usepackage{array}
\usepackage{booktabs}
\usepackage{grffile}
\usepackage{ifthen}
\usepackage{xspace}
\usepackage{subfig}
\usepackage[colorlinks=true,linkcolor=black,urlcolor=black,citecolor=black]
{hyperref}

\def\Title#1{\begin{center} {\Large #1 } \end{center}}
\def\Author#1{\begin{center}{ \sc #1} \end{center}}
\def\Address#1{\begin{center}{ \it #1} \end{center}}

\newcommand\pubblock{\rightline{\begin{tabular}{l} Proceedings of the Second Annual LHCP\\ \pubnumber\\
         \pubdate  \end{tabular}}}

\newenvironment{Abstract}{\begin{quotation} \begin{center} 
             \large ABSTRACT \end{center}\bigskip 
      \begin{center}\begin{large}}{\end{large}\end{center} \end{quotation}}

\newenvironment{Presented}{\begin{quotation} \begin{center} 
             PRESENTED AT\end{center}\bigskip 
      \begin{center}\begin{large}}{\end{large}\end{center} \end{quotation}}

\def\beq{\begin{equation}}
\def\eeq#1{\label{#1}\end{equation}}
\def\eeqn{\end{equation}}

\def\beqa{\begin{eqnarray}}
\def\eeqa#1{\label{#1}\end{eqnarray}}
\def\eeqan{\end{eqnarray}}

\let\bar=\overbar

\def\Dslash{\not{\hbox{\kern-4pt $D$}}}
\def\dslash{\not{\hbox{\kern-2pt $\del$}}}

\def\msb{{\bar{\ssstyle M \kern -1pt S}}}

\newcommand{\tev}{\ensuremath{\mathrm{\,Te\kern -0.1em V}}\xspace}
\newcommand{\gev}{\ensuremath{\mathrm{\,Ge\kern -0.1em V}}\xspace}

\newcommand{\psec}{\ensuremath{\,\mathrm{ps}}\xspace}
\newcommand{\mm}{\ensuremath{\,\mathrm{mm}}\xspace}
\newcommand{\pt}{\ensuremath{p_\mathrm{T}}\xspace}
\newcolumntype{M}{>{\centering\arraybackslash}m}

\DeclareSubrefFormat{proc}{#2}
\newcommand{\subfig}[2]{\subfloat[]{\includegraphics[width=#1\columnwidth]
    {#2}\label{fig:#2}}}
\newcommand{\citeepos}{\cite{Pierog:2009zt}\xspace}
\newcommand{\citeqgsjet}{\cite{Ostapchenko:2007qb}\xspace}
\newcommand{\citesybill}{\cite{Ahn:2009wx}\xspace}
\newcommand{\citepythia}{\cite{Sjostrand:2006za,Sjostrand:2007gs}\xspace}

\newcommand{\citeherwigpp}{\cite{Bahr:2008pv}\xspace}
\newcommand{\tabref}[1]{Table~\ref{tab:#1}}
\newcommand{\figref}[1]{Fig.~\ref{fig:#1}}

\newcommand\pubnumber{LHCb-PROC-2014-037}
\newcommand\pubdate{\today}
\def\affiliation{
On behalf of the LHCb Collaboration, \\
Laboratory for Nuclear Science \\
Massachusetts Institute of Technology, Cambridge, MA, U.S.A.}

\begin{document}

\large
\begin{titlepage}
\pubblock
\vfill
\Title{Soft QCD Measurements at LHCb}
\vfill
\Author{Philip Ilten}
\Address{\affiliation}
\vfill

\begin{Abstract}
Studies in the forward region of charged particle multiplicity and density, as well as energy flow, are presented. These measurements are performed using data from proton-proton collisions at a center-of-mass energy of $7\tev$, collected with the LHCb detector. The results are compared to predictions from a variety of Monte Carlo event generators and are used to test underlying event and hadronization models as well as the performance of event generator tunes in the forward region.
\end{Abstract}
\vfill

\begin{Presented}
The Second Annual Conference\\
 on Large Hadron Collider Physics \\
Columbia University, New York, U.S.A \\ 
June 2-7, 2014
\end{Presented}
\vfill
\end{titlepage}
\def\thefootnote{\fnsymbol{footnote}}
\setcounter{footnote}{0}
\normalsize 

\section{Introduction}

Within quantum chromodynamics (QCD), calculations can be performed with a perturbative expansion when $\alpha_s$ is small (hard QCD), but must be performed using non-perturbative methods when $\alpha_s$ becomes large (soft QCD). In Monte Carlo (MC) event generators a typical event is factorized into steps: the hard process calculation, initial state radiation (ISR), final state radiation (FSR), underlying event, and hadronization. In this way the hard QCD regime of the first three steps can be separated from the soft QCD regime of the final two steps to provide a complete event description. However, the underlying event and hadronization are not predicted via perturbative QCD and must be tuned with experimental data.

When tuning event generators it is useful to disentangle the effects of the underlying event and hadronization. During hadronization the partons produced from all previous steps are combined into hadrons to create final state particles that can be observed either directly or through subsequent decays. Hadronization can depend upon the ISR/FSR cutoff scale, and so these models can be tuned with infrared sensitive variables such as charged particle multiplicity and density. In hadron collisions the underlying event is oftentimes produced through multi-parton interactions (MPI), where the incoming hadrons are treated as collections of partons and multiple soft interactions, in addition to the hard process, occur. MPI does not depend upon the ISR/FSR cutoff scale and can be tuned with infrared safe observables such as energy flow.

The LHCb detector~\cite{Alves:2008zz} is a fully instrumented single-arm forward spectrometer on the LHC covering the pseudo\-rapidity range $2 \leq \eta \leq 5$ with additional backwards coverage from a vertex locator, and provides soft QCD measurements that not only supply new tuning data, but are also important for validating centrally tuned MPI and hadronization models in the forward region. A summary of soft QCD measurements made with LHCb is given in \tabref{summary} where the center-of-mass energy, {\sc Rivet}~\cite{Buckley:2010ar} analysis plugin number, and reference are given. Results from the first two italicized analyses on charged particle multiplicities~\cite{Aaij:2014pza} and energy flow~\cite{Aaij:2012pda}, are presented in the remainder of this proceeding.

\begin{table}
  \begin{center}
    \begin{tabular}{l|l|l|l}
      \toprule
      \multicolumn{1}{c|}{measurement} 
      & \multicolumn{1}{c|}{$\sqrt{s}\,[\mathrm{TeV}]$} 
      & \multicolumn{1}{c|}{{\sc Rivet} plugin} & \multicolumn{1}{c}{ref.} \\
      \midrule

      {\it charged particle multiplicities and densities}
      & 7
      &  to be released
      & \cite{Aaij:2014pza}
      \\[0.1cm]

      {\it energy flow}
      & 7
      &  LHCB\_2013\_I1208105
      &  \cite{Aaij:2012pda}
      \\[0.1cm]

      prompt charm cross-sections
      & 7
      & LHCB\_2013\_I1218996
      & \cite{Aaij:2013mga}
      \\[0.1cm]

      charged particle ratios
      & 0.9
      & LHCB\_2012\_I1119400
      & \cite{Aaij:2012ut}
      \\[0.1cm]

      $V^0$ ratios
      & 0.9,~7
      & LHCB\_2011\_I917009
      & \cite{Aaij:2011va}
      \\[0.1cm]

      inclusive $\phi$ cross-sections
      & 7
      & LHCB\_2011\_I919315
      & \cite{Aaij:2011uk}
      \\[0.1cm]

      charged particle multiplicities
      & 7
      &
      & \cite{Aaij:2011yj}
      \\[0.1cm]

      prompt $K_S^0$ cross-sections
      & 0.9
      & LHCB\_2010\_S8758301
      & \cite{Aaij:2010nx}
      \\[0.1cm]

      inclusive $b\bar{b}$ cross-sections
      & 7
      & LHCB\_2010\_I867355
      & \cite{Aaij:2010gn}
      \\

    \bottomrule
    \end{tabular}
    \caption{\label{tab:summary}Summary of forward soft QCD measurements made with LHCb. Results from the first two italicized analyses are presented in this proceeding.}
  \end{center}
\end{table}

\section{Charged Particle Multiplicity}

Two hadronization models are primarily used by event generators: the string model~\cite{Andersson:1983ia} and the cluster model~\cite{Fox:1979ag}. The string model is based on the principle of linear confinement where the potential between quarks at large length scales is thought to arise from gluon self-interaction and is roughly linear; quarks are split according to this potential to produce hadrons, resulting in well modeled kinematics but a poorly predicted final state flavor description. In contrast, the cluster model utilizes pre-confinement where proto-hadrons, independent of the hard process scale, are formed and then evolved via two-body decays into final state hadrons. Here, the hadron kinematics are not as well modeled, while the flavor description is.

The LHCb charged particle multiplicity and density measurements~\cite{Aaij:2014pza}, performed using proton-proton collision data at ${\sqrt{s} = 7\tev}$ with a low interaction rate, provide an excellent test of these two models. Here, visible events at the generator level are required to contain at least one charged particle within the pseudo\-rapidity range ${2.0 < \eta < 4.8}$ with a transverse momentum of ${\pt > 0.2\gev}$, a momentum of ${p > 2\gev}$, and a lifetime of ${\tau < 10\psec}$. A reconstructed event must contain at least one track traversing all LHCb tracking stations as well as passing within ${2\mm}$ of the beam-line and originating from the luminous region of the collision.

In order to compare the observed data to MC predictions the reconstructed sample is first corrected for impurities by re-weighting each track on an event-by-event basis with a purity factor. Within the sample ${\approx 6.5\%}$ of tracks are from reconstruction artifacts, ${\approx 1\%}$ are duplicates, and ${\approx 4.5\%}$ are non-prompt. Next, the sample is corrected with the probability for a visible event to occur with no reconstructed tracks. Finally, the distributions from the sample are unfolded for pile-up effects to obtain measurements for single proton-proton collisions and then corrected for reconstruction efficiencies.

\begin{figure}
  \begin{center}
    \begin{tabular}{M{0.45\columnwidth}M{0.45\columnwidth}}
      \subfig{0.45}{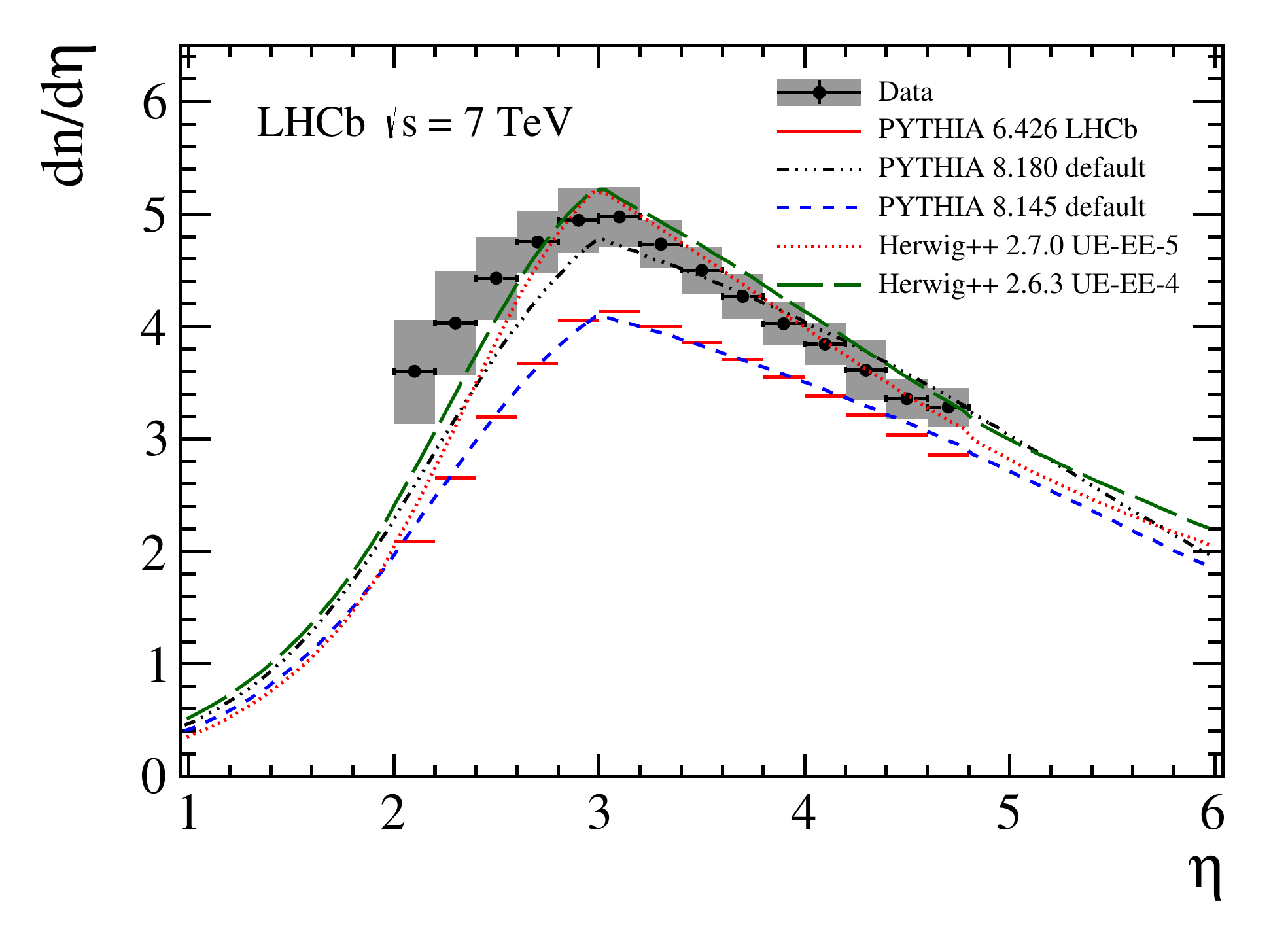} & 
      \subfig{0.45}{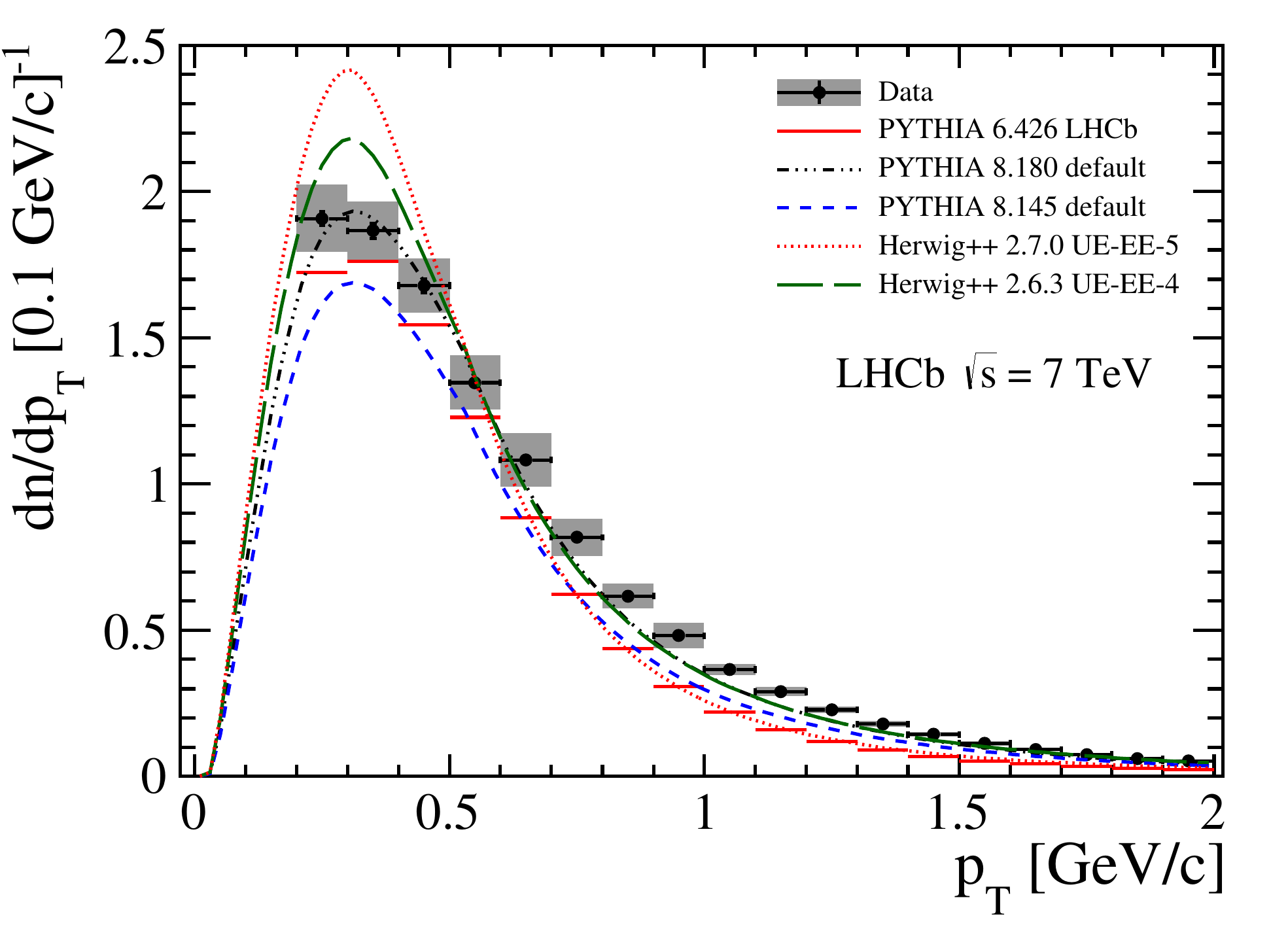} \\
    \end{tabular}
    \vspace{-0.2cm}
    \caption{\label{fig:density}Differential charged particle density as a function of \protect\subref*{fig:eta.dn} pseudo\-rapidity and \protect\subref*{fig:pt.dn} \pt.}
  \end{center}
\end{figure}

\begin{figure}
  \begin{center}
    \begin{tabular}{M{0.45\columnwidth}M{0.45\columnwidth}}
      \subfig{0.45}{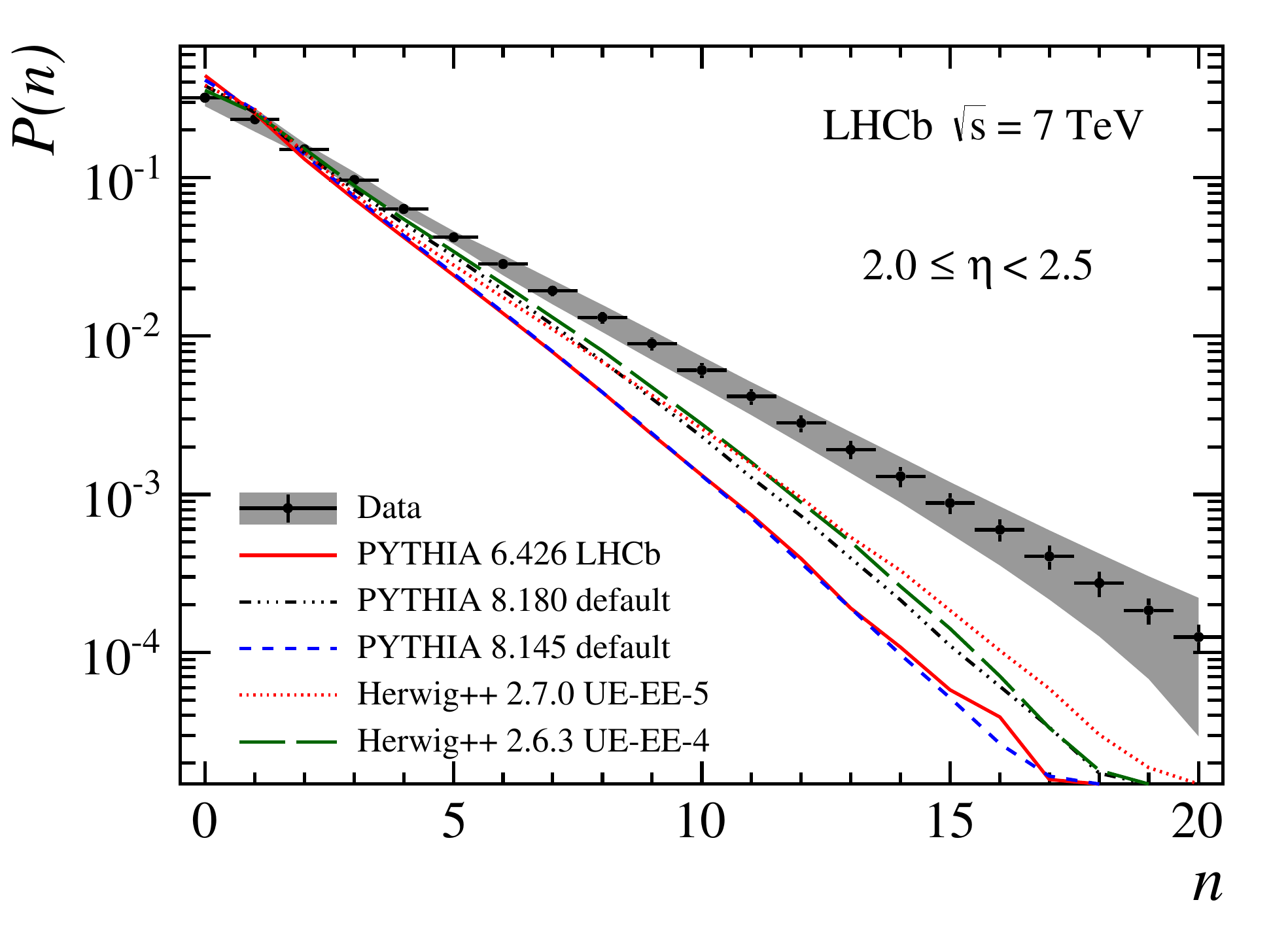} &
      \subfig{0.45}{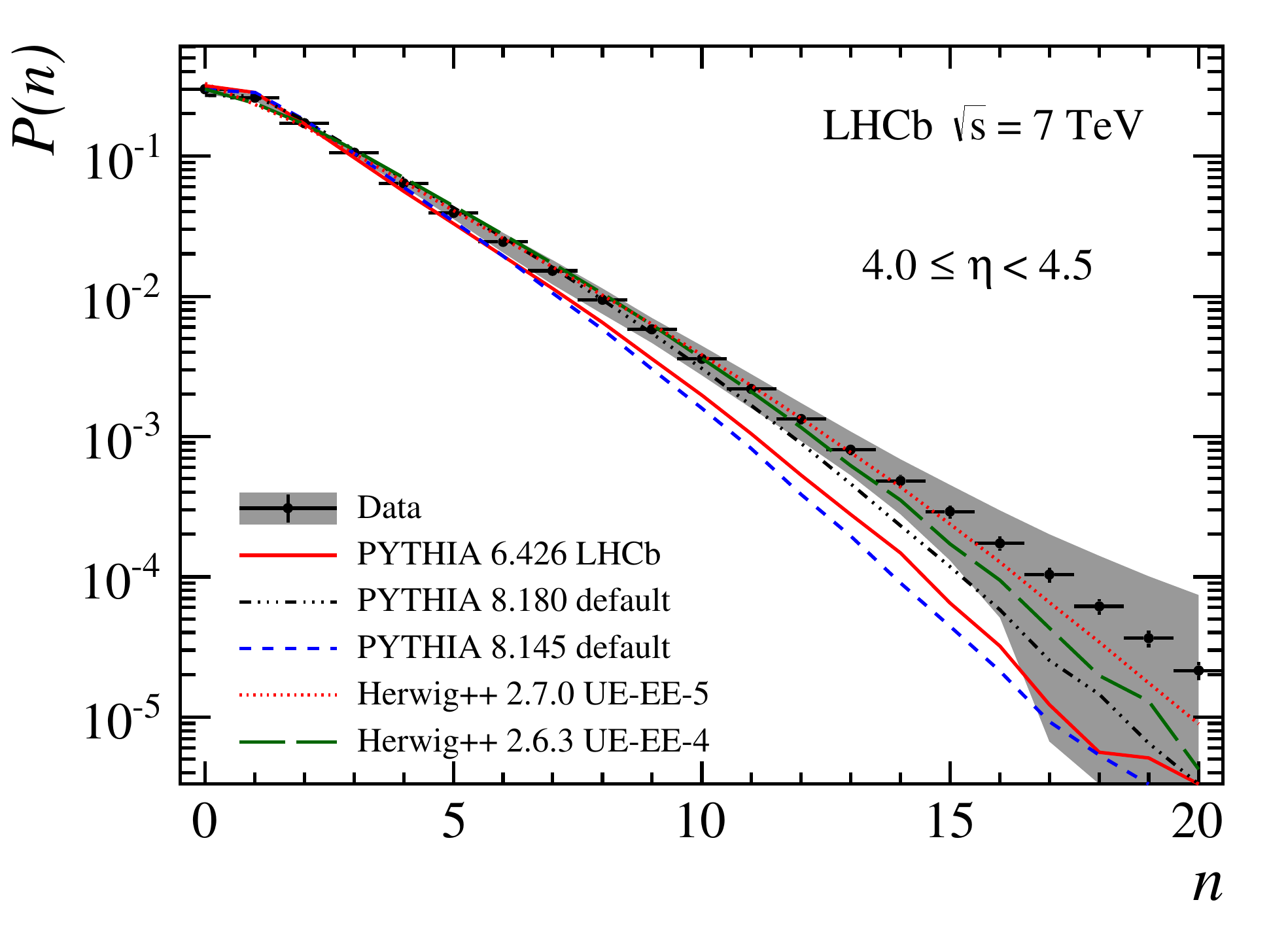} \\
    \end{tabular}
    \vspace{-0.2cm}
    \caption{\label{fig:mult}Charged particle multiplicity for the pseudo\-rapidity ranges \protect\subref*{fig:n.pn.eta.1} ${2.0 < \eta < 2.5}$ and \protect\subref*{fig:n.pn.eta.5} ${4.0 < \eta < 4.5}$.}
  \end{center}
\end{figure}

The results are compared to predictions from {\sc Pythia}~\citepythia which uses a string hadronization model and {\sc Herwig++}~\citeherwigpp which uses a cluster hadronization model. Three tunes with three unique versions of {\sc Pythia} are used: the LHCb tune~\cite{Belyaev:2010yga} with $6.426$, the default $8.145$ tune, and tune-4C~\cite{Corke:2010yf} with $8.180$. For {\sc Herwig++} two tunes~\cite{Gieseke:2012ft} with two unique versions are used: UE-EE-4-MRST with $2.6.3$ and UE-EE-5-MRST with $2.7.0$. All tunes except the first two {\sc Pythia} tunes use initial central LHC data.

In \figref{density} the differential charged particle density as a function of pseudo\-rapidity and \pt is compared to MC predictions. The falling distribution at low $\eta$ is purely due to the momentum requirement. Both the non-LHC tunes significantly under\-estimate the charge density while the remaining tunes describe the data well. However, at low \pt the two {\sc Herwig++} predictions over\-estimate the charge density, with the UE-EE-4-MRST tune providing a slightly more consistent description of the data. Tune-4C provides an excellent description of the data in both $\eta$ and $\pt$. The charged particle multiplicity is plotted in \figref{mult} in the pseudo\-rapidity bins ${2.0 < \eta < 2.5}$ and ${4.0 < \eta < 4.5}$. At low $\eta$ all the tunes significantly under\-estimate high charge multiplicities, with the LHC tunes performing slightly better. At high $\eta$ the LHC tune predictions for both {\sc Pythia} and {\sc Herwig++} agree well with data, while the non-LHC tunes do not. Further comparisons in bins of $\eta$ and \pt from~\cite{Aaij:2014pza} demonstrate none of the models fully describe the data.

\section{Energy Flow}

Multi-parton interactions can be modeled going from hard $\to$ soft QCD, as done by high energy physics generators, or from soft $\to$ hard, as done by many air-shower generators. In the hard $\to$ soft method $2 \to 2$ $t$-channel QCD exchanges are generated with the low \pt divergence regulated with either a cutoff or damping parameter, which models color screening and saturation effects. The soft $\to$ hard models begin with the exchange of color-singlet pomerons which are resolved into hard $gg$ structure at higher energies through a smooth transition.

The LHCb energy flow measurements~\cite{Aaij:2012pda}, also using proton-proton collision data at ${\sqrt{s} = 7\tev}$ with a low interaction rate, allow these two models to be tested in the forward region. The charged energy flow is defined in bins of pseudo\-rapidity as,
\begin{equation}
  \frac{1}{N_\mathrm{int}} \frac{\mathrm{d}E}{\mathrm{d}\eta}
  = \frac{1}{\Delta \eta} \left(\frac{1}{N_\mathrm{int}} \sum_{i =
    1}^{N_\mathrm{trk}(\Delta \eta)} E_i \right)
\end{equation}
where $N_\mathrm{int}$ is the number of inelastic proton-proton interactions, $N_\mathrm{trk}$ is the number of tracks in the bin range $\Delta \eta$, and $E_i$ is the energy of each track. Only events with zero or one primary vertices are used, and all tracks are required to have a momentum within ${2 < p < 1000\gev}$. The inclusive data sample is sub-divided into events with,
\begin{itemize}
  \item {\it hard-scattering}: $\geq 1$ tracks with ${\pt > 3\gev}$ and ${1.9 < \eta < 4.9}$
  \item {\it diffractive}: $< 1$ tracks within ${-3.5 < \eta < -1.5}$
  \item {\it non-diffractive}: $\geq 1$ tracks within ${-3.5 < \eta < -1.5}$
\end{itemize}
where the {\it diffractive} definition corresponds to an $\approx 70\%$ purity as estimated with the {\sc Pythia 6} model and the {\it non-diffractive} definition corresponds to an $\approx 90\%$ purity. The measured distributions are unfolded for detector effects using bin-to-bin corrections, with a systematic uncertainty estimated from various {\sc Pythia} configurations. The total energy flow is calculated by estimating the neutral energy flow from the charged energy flow, using simulation from {\sc Pythia} 6 and the neutral to charged ratio determined from data.

\begin{figure}
  \begin{center}
    \begin{tabular}{M{0.33\columnwidth}M{0.33\columnwidth}}
      \subfig{0.33}{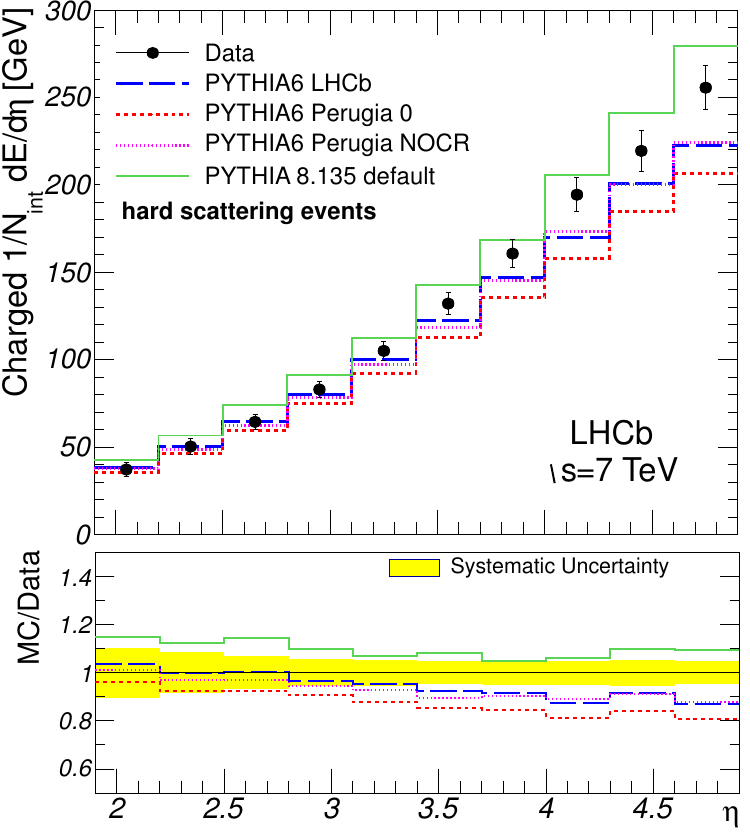} & 
      \subfig{0.33}{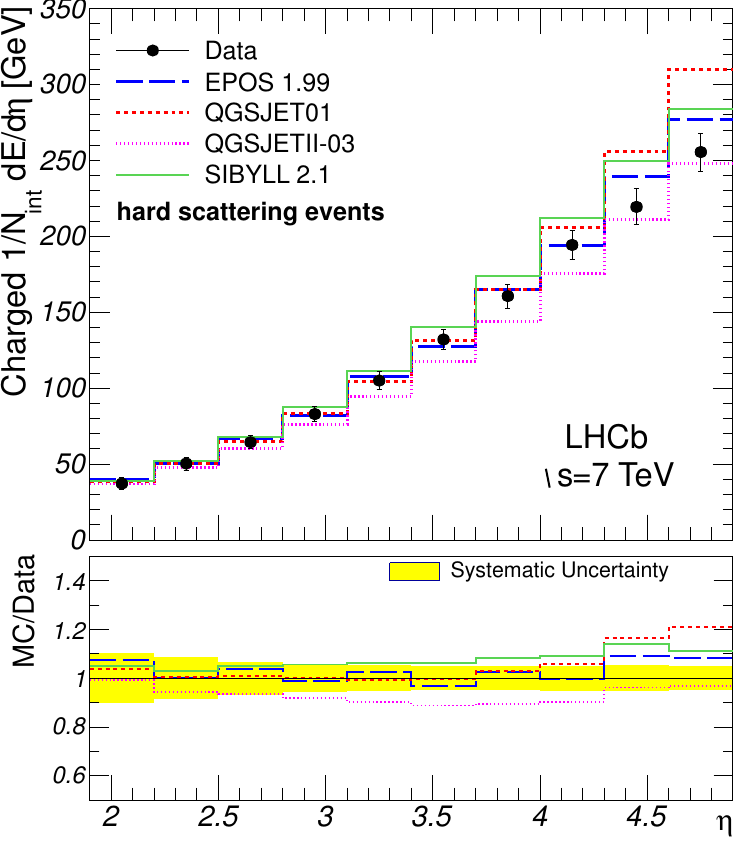} \\
      \subfig{0.33}{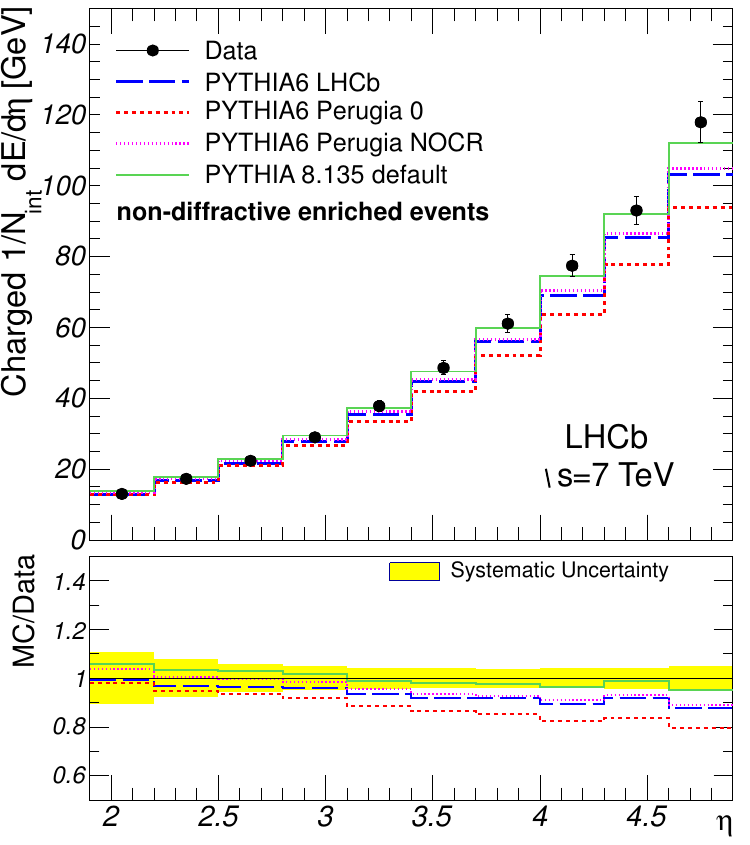} & 
      \subfig{0.33}{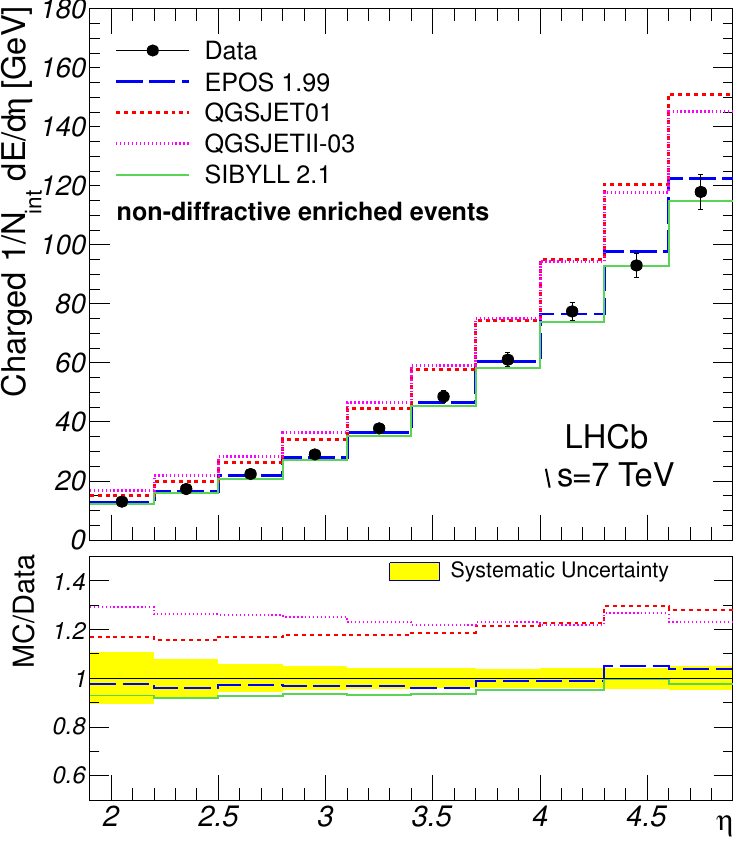} \\
      \subfig{0.33}{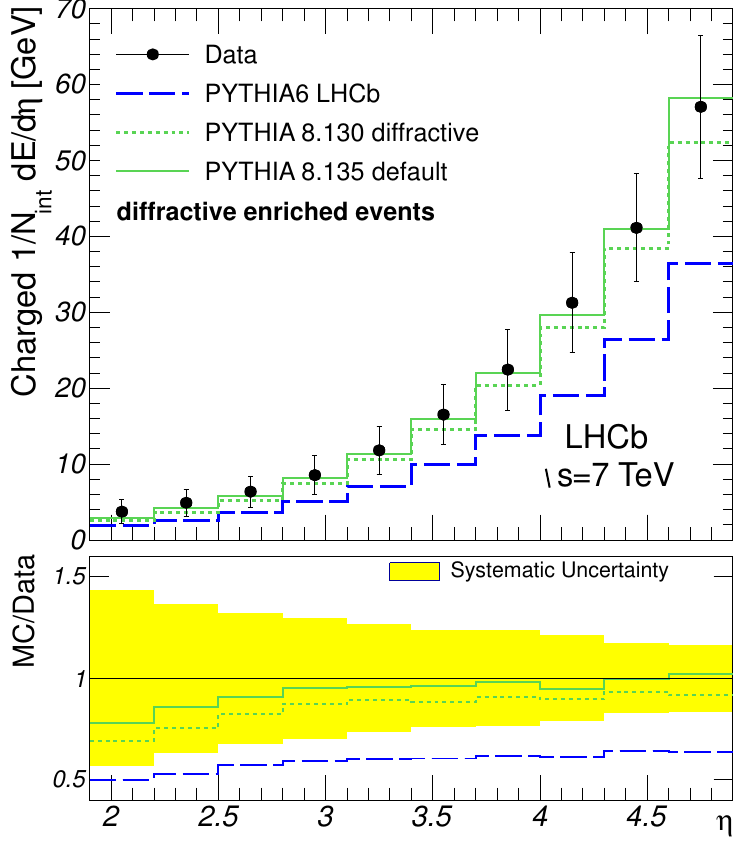} & 
      \subfig{0.33}{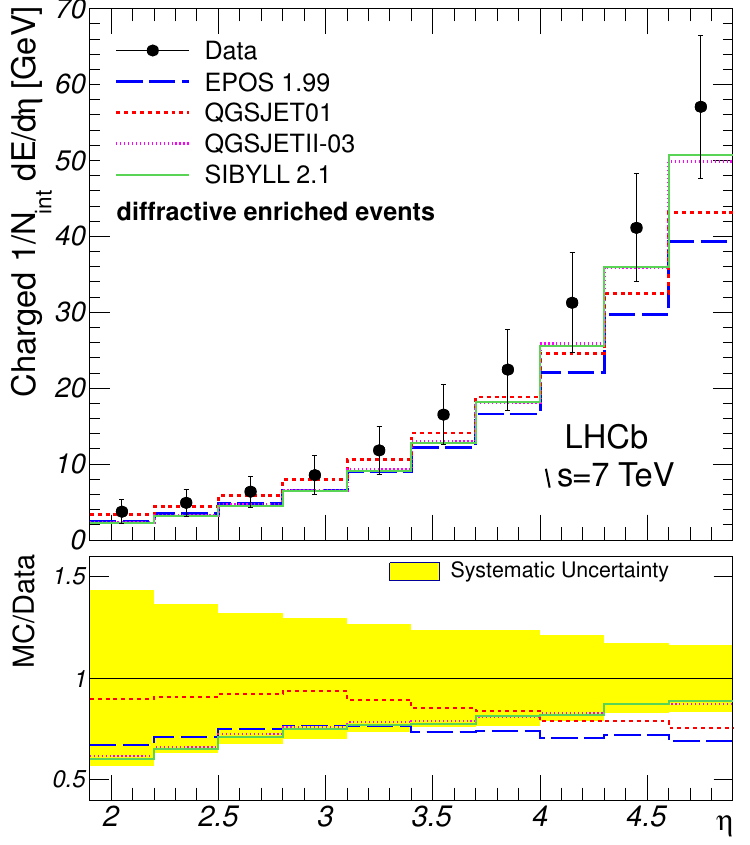} \\
    \end{tabular}
    \caption{\label{fig:eflow}Charged energy flow as a function of pseudo\-rapidity compared with (left) hard $\to$ soft and (right) soft $\to$ hard MPI models. The events are divided into \protect\subref*{fig:eta.echr.hs.1}-\protect\subref*{fig:eta.echr.hs.2} {\it hard-scattering}, \protect\subref*{fig:eta.echr.nd.1}-\protect\subref*{fig:eta.echr.nd.2} {\it non-diffractive}, and \protect\subref*{fig:eta.echr.ad.1}-\protect\subref*{fig:eta.echr.ad.2} {\it diffractive}  enriched samples.}
  \end{center}
\end{figure}

In \figref{eflow} the measured charged energy flow as a function of pseudo\-rapidity is compared to {\sc Pythia} predictions, a hard $\to$ soft MPI model, and air-shower generator predictions, soft $\to$ hard models. Four {\sc Pythia} tunes are provided for comparison: the LHCb tune, the Perugia tune~\cite{Skands:2010ak} with and without color reconnection (NOCR), and the default {\sc Pythia} $8.135$ tune. Predictions from four air-shower generators are provided: {\sc Epos} 1.99~\citeepos, {\sc QgsJet}01 and {\sc QgsJetII-3}~\citeqgsjet, and {\sc Sibyll} 2.1~\citesybill. None of the predictions supplied are based on LHC data. The total energy flow which can be found in~\cite{Aaij:2012pda} is omitted as it exhibits a similar behavior to the charged energy flow.

For {\it hard-scattering} enriched events {\sc Pythia 8} uniformly over\-estimates the data while {\sc Pythia 6} under\-estimates for large $\eta$; the data for middle $\eta$ is under\-estimated by {\sc QgsJetII-03} while the remaining air-shower generators over\-estimate the data for large $\eta$. Note that for large $\eta$ the experimental uncertainty is smallest, while the differences between models is largest. The {\it non-diffractive} enriched data is well modeled by {\sc Pythia 8}, {\sc Pythia 6} with no color reconnection, {\sc Epos}, and {\sc Sibyll}, while {\sc Pythia 6} with color reconnection under\-estimates at large $\eta$ and {\sc QgsJet} uniformly over\-estimates. Finally, the {\it diffractive} enriched data is well described by {\sc Pythia 8} but under\-estimated by all remaining generators.

\section{Conclusions}

LHCb provides complementary soft QCD measurements to the general purpose LHC detectors in the high pseudo\-rapidity and low \pt regions, with {\sc Rivet} plugins available for many. These results validate the consistency of MPI and hadronization models at LHC energies within the forward region with no unexpected behavior observed. The non-LHC tunes under\-estimate forward particle density, while the {\sc Pythia 8} tune-4C consistently describes the data well.

\end{document}